\documentclass[fleqn,10pt]{wlscirep}
\usepackage[utf8]{inputenc}
\usepackage[T1]{fontenc}
\usepackage{amsmath,amsfonts,amssymb}
\usepackage{graphicx}
\usepackage{setspace}
\usepackage{tocloft}
\usepackage{epsfig}
\usepackage{subfig}
\usepackage{graphicx}
\usepackage{amsmath}
\usepackage{array}
\usepackage{here}
\usepackage{bm}
\usepackage{comment}
\usepackage{soul}
\graphicspath{images/}
\title{Response of atomic spin-based sensors to magnetic and nonmagnetic perturbations}

\author[1,*]{Mikhail Padniuk}
\author[1]{Marek Kopciuch}
\author[2,3]{Riccardo Cipolletti}
\author[2]{Arne Wickenbrock}
\author[2]{Dmitry Budker}
\author[1,+]{Szymon Pustelny}

\affil[1]{Marian Smoluchowski Institute of Physics, Jagiellonian University, {\L}ojasiewicza 11, 30-348, Krak{\'o}w, Poland.}
\affil[2]{Helmholtz Institute, Johannes Gutenberg-Universitat at Mainz, 55099 Mainz, Germany}
\affil[3]{Robert Bosch GmbH, Corporate Sector Research and Advance Engineering, Advanced Technologies and Micro Systems, 71272 Renningen, Germany}
\affil[*]{michal.padniuk@doctoral.uj.edu.pl}
\affil[+]{szymon.pustelny@uj.edu.pl}

\newcommand{\old}[1]{}
\newcommand{\new}[1]{{{{#1}}}}

\begin{abstract}
    Searches for pseudo-magnetic spin couplings require implementation of techniques capable of sensitive detection of such interactions. While Spin-Exchange Relaxation Free (SERF) magnetometry is one of the most powerful approaches enabling the searches, it suffers from a strong magnetic coupling, deteriorating the pseudo-magnetic coupling sensitivity. To address this problem, here, we compare, via numerical simulations, the performance of SERF magnetometer and noble-gas-alkali-metal co-magnetometer, operating in a so-called self-compensating regime.  We demonstrate that the co-magnetometer allows reduction of the sensitivity to low-frequency magnetic fields without loss of the sensitivity to nonmagnetic couplings. Based on that we investigate the responses of both systems to the oscillating and transient spin perturbations. Our simulations reveal about five orders of magnitude stronger response to the neutron pseudo-magnetic coupling and about three orders of magnitude stronger response to the proton pseudo-magnetic coupling of the co-magnetometer than those of the SERF magnetometer. Different frequency responses of the co-magnetometer to magnetic and nonmagnetic perturbations enables differentiation between these two types of interactions.
    This outlines the ability to implement the co-magnetometer as an advanced sensor for the Global Network of Optical Magnetometer for Exotic Physics searches (GNOME), aiming at detection of ultra-light bosons (e.g., axion-like particles).
\end{abstract}
\begin{document}

\flushbottom
\maketitle
%
%
\thispagestyle{empty}
\section*{Introduction}

\noindent  In optically pumped magnetometers (OPMs), measurements of external magnetic fields rely on detection of energy-level shifts/spin precession arising from Zeeman interaction \cite{budker2013optical}.  However, if nonmagnetic interactions similarly affect the energy levels, OPMs can  also be used to detect those interactions.  This enables the application of optical magnetometry to searches for anomalous spin-dependent interactions.  

Currently, OPMs are used to search for physics beyond the Standard Model in a variety of experiments (see, Ref.~\cite{Safronova2018Search} and references therein).  A particular example of such OPMs application is the search for microscopic-range spin-dependent interactions, indicating a possibility of existence of axion-like particles (ALPs), which are one of prime candidates for the dark matter \cite{Safronova2018Search}.  OPMs are also used to search for transient nonmagnetic spin couplings, which could arise due to interaction with macroscopic objects made of ALPs, in particular, Q-balls \cite{Kimball2018Searching}, topological defects (e.g., domain walls) of ALP field \cite{Pospelov2013Detecting}, or ALP-field pulses generated in cataclysmic astrophysical events (e.g., black-hole mergers) \cite{Dailey2021Quantum}.  These transient couplings are targeted by the Global Network of Optical Magnetometers for Exotic physics searches (GNOME) \cite{Pustelny2013Global,Afach2018Characterization, afach2021search}.  Heretofore, the GNOME consists of various OPMs, originally developed for ultra-sensitive magnetometry in globally distributed locations.  This leads to several challenges  when searching the data for global transient signals.  First, due to a different nuclear-spin content of atoms used in specific sensors, they are characterised with different sensitivity to exotic spin couplings\cite{Kimball_2015} (coupling to protons and neutrons could be, in general, different).  Second, the implemented OPMs are characterised with different bandwidths, sensitivities, and  local noise floors, which complicates data analysis\cite{MASIAROIG2020_analysis_method}.  Third, the magnetometers were designed to maximise the sensitivity to magnetic fields. Uncontrolled and uncompensated magnetic-field perturbations are detrimental to the sensitivity to other couplings.  These issues triggered work to upgrade conventional OPMs in the GNOME with a sensor less sensitive to magnetic fields but highly sensitive to nonmagnetic spin couplings. 

A specific example of a sensor, being predominantly sensitive to nonmagnetic spin couplings, hence well suited for searches for physics beyond the Standard Model, is an alkali-metal-noble-gas co-magnetometer originally developed by Romalis and coworkers \cite{Romalis2002Dynamic, Kornack_Nuclear_spin_gyro,Ghosh_Romalis_SEOP_Ne} and later studied extensively by other researchers \cite{Rot_sens_Li,fang2016low, chen2016spin, Co-mag_magn_field_resp_Fan, Shi:20}.  Such a system operates based on coupled evolution of the magnetizations of noble gas (NG) and alkali-metal (AM) vapour.  Both can achieve a high percentage of polarisation as AM can be optically pumped, whereas polarisation of the NG can be generated via spin-exchange collisions with the optically polarised AM \cite{theory_of_he_SEOP}.  In the co-magnetometer, the vapour cell is heated above 150$^\circ$C to achieve a sufficiently high AM density, such that relaxation due to spin-exchange collisions, which is one of the main mechanisms of AM polarisation relaxation and hence one of a limiting factor to spin-coupling sensitivity, is suppressed. This is the so-called Spin-Exchange Relaxation Free (SERF) regime \cite{allred2002high}. Furthermore, for the alkali-metal-noble-gas co-magnetometer the effect of low-frequency magnetic drifts can be suppressed by application of  a carefully chosen bias magnetic field. If the bias field is approximately equal to the sum of AM and NG magnetization fields, the system retains high sensitivity to both electron and nuclear nonmagnetic spin couplings, but becomes insensitive to low-frequency magnetic-field changes. 
Such self-compensating co-magnetometers have already been  used for tests of the Lorentz symmetry \cite{Kornack_phdthesis, new_CPT_limit,New_test_of_local_lorentz_invariance}, setting limit on the neutron coupling to light pseudoscalar particles \cite{Limits_on_new_long_range_nuclear_spin_dep_forces_Romaslis}, and spin-mass interaction of  fermions \cite{New_limit_spin_mass_lee}. In all of those applications, however, the signals of interests had low frequencies (typically below a few Hz), for which the response to  magnetic fields is almost entirely suppressed and the system is only sensitive  to nonmagnetic spin couplings.  Since the GNOME targets transient signals, a question of the co-magnetometer's applicability  to such searches is important and well motivated. Additional interest in co-magnetometer systems, in the context of searches for transient effects, arises from the possibility for quantitative distinguishing between magnetic and nonmagnetic transients. This was originally considered for precise rotation sensing with NG-AM co-magnetometers\cite{Kornack_Nuclear_spin_gyro} but can be extended for other scenarios. 

In this work, we analyse the alkali-metal-noble-gas co-magnetometer in the context of its response to time-dependent electron and nuclear spin perturbations, and we compare the results with the response of the AM SERF magnetometer. Our theoretical analysis is based on numerical solution of differential equations describing the coupled evolution of the AM and NG.  We use Ordinary Differential Equations (ODE) to simulate the response  of atoms to  spin perturbations in the absence of noise and we provide analytical models, which reproduce numerical results.  The discussions concern the responses of both, SERF magnetometer and co-magnetometer, to magnetic fields and pseudo-magnetic spin perturbations, where in the latter case we independently consider the effects of electron,  proton, and neutron spin perturbations.  These allow us to determine the frequency and phase responses of both devices. Finally, we compare the response of the co-magnetometer and SERF magnetometer to transient effects of both magnetic and nonmagnetic nature. We show that differences between the co-magnetometer frequency responses for different spin perturbations allows identification of nonmagnetic transient effects even with  a single sensor. These additional signatures of the nonmagnetic transient effects can be utilised to decrease the false-positive event rate in searches with the GNOME network. 

\section*{Methods}

\subsection*{Numerical models}
\label{sect:NM}  
In this section, we describe the theoretical models used to simulate the response of the SERF magnetometers and co-magnetometers to magnetic and nonmagnetic spin perturbations.

\subsubsection*{SERF magnetometer model}
In a conventional SERF magnetometer, a spin-zero NG can be used as a buffer gas.  The buffer gas limits diffusion of the atoms towards the cell walls (the AM atoms are depolarised in wall collisions), which increases the polarisation lifetime but, it does not produce any magnetisation. Thereby, to simulate a response of such a magnetometer to a spin perturbation, we implement an approach based on the solutions of the Bloch equation\cite{Romalis_Savukov_2005} with inclusion of nonmagnetic spin couplings (for more details see Additional information). The equation describing the AM polarisation $\mathbf{P}$ of atoms subjected to an external magnetic field $\mathbf{B}$ and circularly polarised light can be written as
\begin{equation}
    \label{eq:BE}
    \frac{d \mathbf{P}}{d t} = \frac{1}{q}\bigg[\gamma_e \mathbf{(B+b_e)\times P}+(q-1)\gamma_e \mathbf{b_N^{AM}\times P}+\mathbf{(s-P)}R_p - {R_c^e}\mathbf{P}\bigg],
\end{equation}
where $\gamma_e$ is the electron gyromagnetic ratio,  $\mathbf{s}$ is the optical pumping vector, $R_p$ is the pumping rate, $R_c^e$ is the polarisation-relaxation rate, $q$ is the slowing-down factor, which is a function of  the nuclear spin of the AM and its polarisation $\mathrm{P}$. The vectors $\mathbf{b_e}$ and $\mathbf{b_N^{AM}}$ are the nonmagnetic electron and nuclear spin perturbations, respectively, given in the magnetic units (pseudo-magnetic field).

\subsubsection*{Co-magnetometer model}
\label{subsec:Comag_mum_model}
In the co-magnetometer, a polarised AM and NG (in this case NG with nonzero nuclear spin is used, so its nuclei can be polarised) occupy the same volume inside of a spherical glass cell. Then, the response of the co-magnetometer to magnetic and nonmagnetic perturbations is determined  by a set of coupled Bloch equations (the so-called Bloch-Hasegawa equations) \cite{Romalis2002Dynamic} with inclusion of nonmagnetic spin perturbations (for more details see Additional information)
\begin{equation}
    \begin{cases}
        \frac{d \mathbf{P^e}}{d t} &= \frac{1}{q}\bigg[\gamma_e \mathbf{(B+b_e+}\lambda M^n\mathbf{ P^n)\times P^e}
        +(q-1)\gamma_e \mathbf{b_N^{AM}\times P^e}+R_{se}^{ne}\mathbf{P^n} +{\mathbf{(s-P^e)}}R_p- ({R_{c}^e}+R_{se}^{en})\mathbf{P^e} \bigg], \\
            \frac{d \mathbf{P^n}}{d t}& =\gamma_n \mathbf{(B+b_N^{NG}+}\lambda M^e\mathbf{ P^e)\times P^n}+ R_{se}^{en}\mathbf{P^e}- (R_{se}^{ne}+R_{c}^n)\mathbf{P^n},
    \end{cases}
\label{eq:BHE_1}
\end{equation}
where $\mathbf{P^e}$ and $\mathbf{P^n}$ stand for the electron polarisation of the AM  and nuclear polarisation of the NG, respectively, $\lambda$ is the coupling-strength factor for interaction between the two polarisations \cite{Schaefer_freq_shifts}, $M^e$ and $M^n$ are the maximal possible magnetisations of AM and NG, and $\gamma_n$ is the nuclear gyromagnetic ratio of NG, $R_{c}^e$ and $R_{c}^n$ are the electron and nuclear polarisation-relaxation rates which take into consideration relaxation due to various interatomic spin destruction collisions and collisions with a walls of a cell. $R_{se}^{ne}$ and $R_{se}^{ne}$ are rates of the polarisation transfer in spin exchange collisions form NG to AM and from AM to NG, respectively. The nonmagnetic nuclear perturbation of the NG spins is denoted by $\mathbf{b_N^{NG}}$.

In this work, we consider the system dynamics over a time scale much shorter than the characteristic time of NG-AM spin-exchange collisions, $t\ll (R_{se}^{ne})^{-1}$.  Thereby, in Eq.~\eqref{eq:BHE_1}, we neglect the term $R_{se}^{ne}\mathbf{P^n}$, which characterises the back action of NG on AM. In fact, in our description we use the steady-state polarisations of AM $\mathbf{P_0^e}$ and NG $\mathbf{P_0^n}$ with total relaxation rates for both AM and NG ($R^e$ and $R^n$, respectively)
\begin{equation}
    \begin{split}
        R^e &= R_p+R_c^e+R_{se}^{en},\\
        R^n &= R_{se}^{ne}+R_c^n,\\
        \mathbf{P_0^e} &= \frac{R_p}{R^e}\mathbf{s},\\
        \mathbf{P_0^n} &= \frac{R_{se}^{en}}{R^n}\mathbf{P_0^e}. 
    \end{split}
\end{equation}
For simplicity of the model, in the simulations both the steady state polarisations and relaxation rates are assumed to be independent.

With such a parameterisation the Eq.~\eqref{eq:BHE_1} have the following form
\begin{equation}
    \begin{cases}
        \frac{d \mathbf{P^e}}{d t} &= \frac{1}{q}\bigg[\gamma_e \mathbf{(B+b_e+}\lambda M^n\mathbf{ P^n)\times P^e}
        +(q-1)\gamma_e \mathbf{b_N^{AM}\times P^e}+(\mathbf{P_0^e-P^e})R^e \bigg], \\
            \frac{d \mathbf{P^n}}{d t}& =\gamma_n \mathbf{(B+b_N^{NG}+}\lambda M^e\mathbf{ P^e)\times P^n+( P^n_0} - \mathbf{P^n})R^n.
    \end{cases}
\label{eq:BHE}
\end{equation}

In order to fully capitalise on the co-magnetometric capabilities (self-compensation of slow magnetic fields), here we consider the operation in the self-compensating regime, which is achieved when a static magnetic field $\mathbf{B_c}$
\begin{equation}
    \mathbf{B_c =}  -(\lambda M^e P_0^e+\lambda M^n P_0^n)\mathbf{z},
    \label{eq:comp_point}
\end{equation}
is applied to the system\cite{Romalis2002Dynamic} (here we assumed that the initial AM and NG polarisations are oriented along the $\mathbf{z}$ axis).

\subsection*{Nuclear spin content and sensitivity to neutron and proton spin perturbations}
\label{subsection:Nuclear_spin_content}
Due to the composite nature of atomic nuclei, the nuclear response may arise due to coupling to protons, neutrons, or a combination of both. Therefore, the effective pseudo-magnetic fields $\mathbf{b_N^{AM}}$ and $\mathbf{b_N^{NG}}$ can be divided into parts: $\mathbf{b_p}$ affecting the protons and $\mathbf{b_n}$ acting on the neutrons \cite{Kimball_2015}
\begin{equation}
\begin{split}
    \mathbf{b_N^{i}}&=\mathbf{b_n^{i}+b_p^{i}},
\end{split}
\end{equation}
where $i$ may stand for either AM or NG. These pseudo-magnetic fields are determined by the nonmagnetic field $\bm{\Xi}$ and coupling constants $\chi_n$ and $\chi_p$ characterising the coupling to neutrons and protons, respectively
\begin{subequations}
    \begin{align}
    \mathbf{b^{AM}_j} &=-\frac{\sigma^{AM}_j }{\mu_B g_s}\chi_j\bm{\Xi},\\
    \mathbf{b^{NG}_j} &= \frac{ \sigma^{NG}_j}{\mu_Ng_{K}}\chi_j\bm{\Xi},
    \end{align}
    \label{eq:effective_fields_body}
\end{subequations}
where $j$ indicates either proton or neutron, $\sigma_j$ corresponds to the proton or neutron fraction of the nuclear spin polarisation of AM and NG (denoted with upper indices), $\mu_B$ is the Bohr magneton, $\mu_N$ is the nuclear magneton, $g_S$ is the AM Land\'{e} factor, and $g_K$ is the NG nuclear spin $g$-factor.  Equations~\eqref{eq:effective_fields_body} show that the effective pseudo-magnetic fields for the NG  are generally different from those for the AM.  Thus, for the simulations or interpretation of results, it is convenient to introduce scaling factors $\eta_j$  which allows comparison between the response of the SERF magnetometer and the co-magnetometer to nonmagnetic spin couplings of the same strength
\begin{equation}
     \mathbf{b^{AM}_j}= \eta_j \mathbf{b^{NG}_j},
\end{equation}
where the scaling factors are
\begin{equation}
    \eta_j = - \frac{\sigma^{AM}_j}{\sigma^{NG}_j}\frac{\mu_N g_K}{\mu_Bg_S}.
    \label{eq:sacaling_for_the_same_strength}
\end{equation}

In the simulations presented in this paper, we consider the responses of the magnetometers to the perturbation of the same coupling strength. This approach takes into account the scaling factors defined in Eq.~\eqref{eq:sacaling_for_the_same_strength}.  One can find a more detailed discussion of the nonmagnetic spin couplings in the SERF and the co-magnetometer in Additional information. 

\subsection*{Simulation parameters}
\label{subsec:simulation_parameters}
In the case of both, the SERF and co-magnetometer, the spin polarisation is monitored through measurements of the AM-polarisation projection on a given direction (here it is the $\mathbf{x}$ axis). As in both systems the atomic species are initially polarised along the $\mathbf{z}$ axis, both magnetometers are primarily sensitive to perturbations along $\mathbf{y}$ (the sensitive direction is determined by the torque generated by the external fields, rotating the spins around the sensitive direction). Specifically, it can be shown from Eqs. (\ref{eq:BE}) and (\ref{eq:BHE}) that the magnetic or pseudo-magnetic field applied along $\mathbf{y}$ rotates the initial polarisation in the $\mathbf{xz}$ plane, which results in a change of the polarisation projection on the $\mathbf{x}$ axis. At the same time, a field applied along the $\mathbf{x}$ axis generates rotation in the $\mathbf{yz}$ plane, therefore the projection of the polarisation on the $\mathbf{x}$ axis remains unchanged. 

In the simulations, we assume that the SERF magnetometer operates using $^{39}$K atoms and the co-magnetometer operates using a $^{39}$K-$^3$He mixture.  Parameters of potassium vapour are chosen to be exactly the same for both systems, so we can properly compare the sensitivity. The concentration of the alkali metal is equal to $ 10^{14}$ cm$^{-3}$, which corresponds to saturated atomic K vapour at {190$^\circ$C}. The concentration of $^3$He is $10^{20}$ cm$^{-3}$, which corresponds to 3.5 amg.  The assumed total relaxation rates $R^e = $1200\,s$^{-1}$ for potassium and  $R^n =5\cdot 10^{-5}$\,s$^{-1}$ helium, corresponding to a lifetime of about {6 h}, well reproduce the experimental conditions. The steady-state polarisation of the AM is $P^e_0=$0.5, which ensures the highest amplitude of the co-magnetometer response\cite{Kornack_phdthesis}, and also corresponds to typical experimental conditions. Polarisation of the NG is  chosen to be $P^n_0=$0.05, which corresponds to typical experimental conditions. These parameters lead to the compensation-field value of {$-131$~nT}. The other simulation parameters (a complete list) is given in Additional information.

In case of $^{39}$K and $^3$He the relation between the effective magnetic fields for the AM and the NG generated by the same nonmagnetic perturbation defined in Eq. (\ref{eq:sacaling_for_the_same_strength}) leads to  the following scaling factors
\begin{subequations}
\begin{align}
\mathbf{b_p^{^{39}K}} &= \eta_p\mathbf{b_p^{^{3}He}}\approx  10^{-3}\ \mathbf{b_p^{^{3}He}},\\
\mathbf{b_n^{^{39}K}} &= \eta_n \mathbf{b_n^{^{3}He}} \approx  10^{-5}\ \mathbf{b_n^{^{3}He}},
\end{align}
\label{eq:sacaling_for_the_same_strength_num_value}
\end{subequations}
where the nuclear spin content information was taken from Ref.~\cite{Kimball_2015}.  

\section*{Results and discussion}
\subsection*{Frequency response to different spin perturbations}
\label{sect:bandwidths}
In this section, we compare responses of the SERF and co-magnetometer devices to various spin perturbations. We analyse the response of the devices by investigating their signals when perturbed with an either magnetic or nonmagnetic periodic $\mathbf{y}$-oriented field $\mathbf{A}$ 
\begin{equation}
       \mathbf{A} = A_0 \sin(2\pi \nu t)\mathbf{y},
       \label{eq:perturbation}
\end{equation}
where $A_0$ is  the amplitude and $\nu$ is the  frequency of the field. For the simulations, we assume that the amplitude of the perturbation is low enough so that the co-magnetometer continuously operates in the self-compensating regime.
To analyse the response, the simulated data are fitted with the function
\begin{equation}
    S=S_0\sin(2\pi f t +\phi),
    \label{eq:fitmodel}
\end{equation}
where $S_0$, $\phi$ and $f$ are, respectively, the amplitude, phase, and frequency of the fitted signal. To avoid distortions in the fit, we ignore transient phenomena at the beginning of the simulations and just fit the dynamical steady-state data.
 
\begin{figure}[h!]
\centering
\begin{minipage}{0.49\linewidth}
    \textbf{(a)}
    \begin{center}
    \includegraphics[width = 0.9\linewidth]{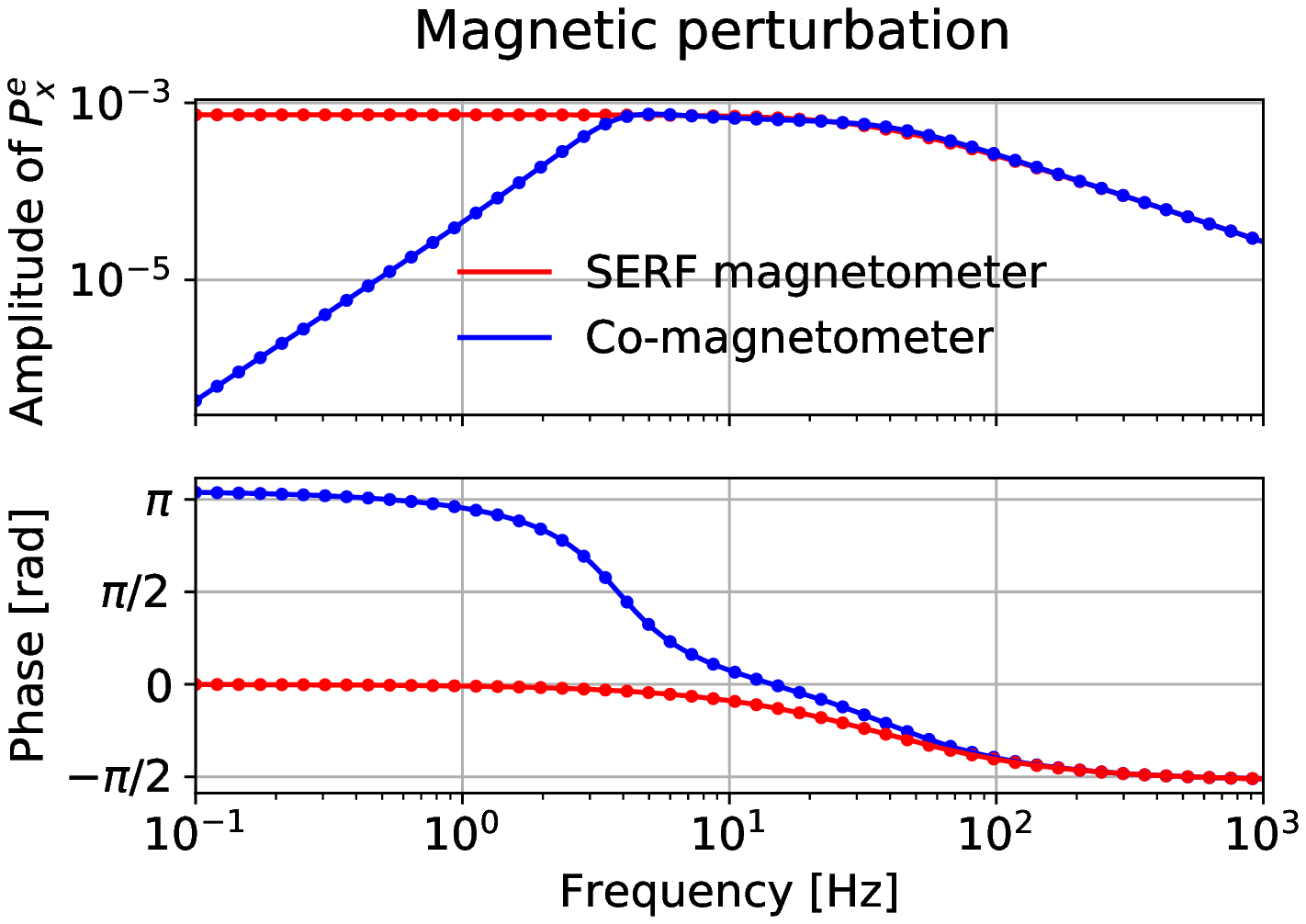}
    \end{center}
\end{minipage}
\hfil
\begin{minipage}{0.49\linewidth}
    \textbf{(b)}
    \begin{center}
    \includegraphics[width = 0.9\linewidth]{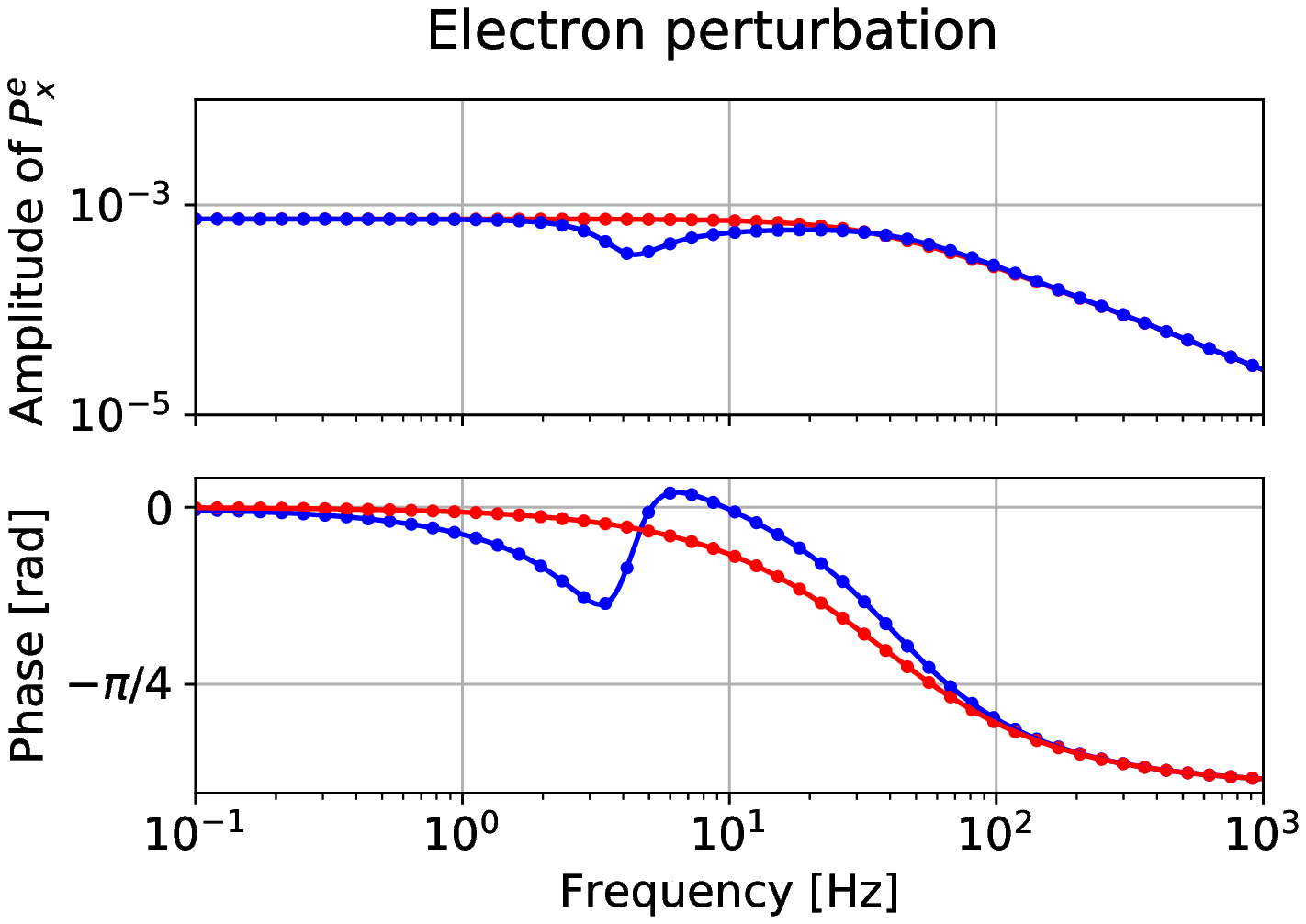}
    \end{center}
\end{minipage}
\hfil
\begin{minipage}{0.49\linewidth}
    \textbf{(c)}
    \begin{center}
    \includegraphics[width = 0.9\linewidth]{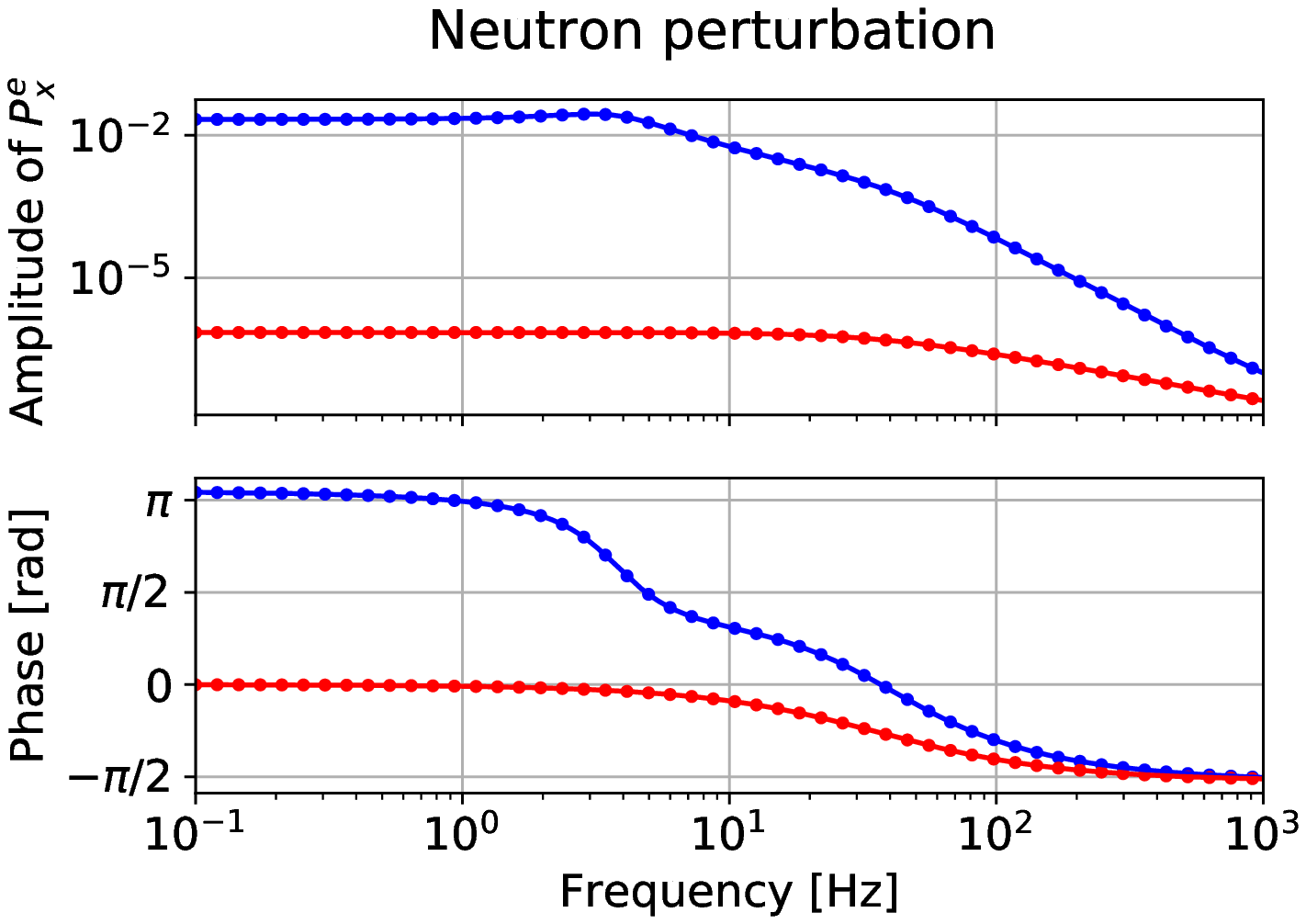}
    \end{center}
\end{minipage}
\hfil
\begin{minipage}{0.49\linewidth}
    \textbf{(d)}
    \begin{center}
    \includegraphics[width = 0.9\linewidth]{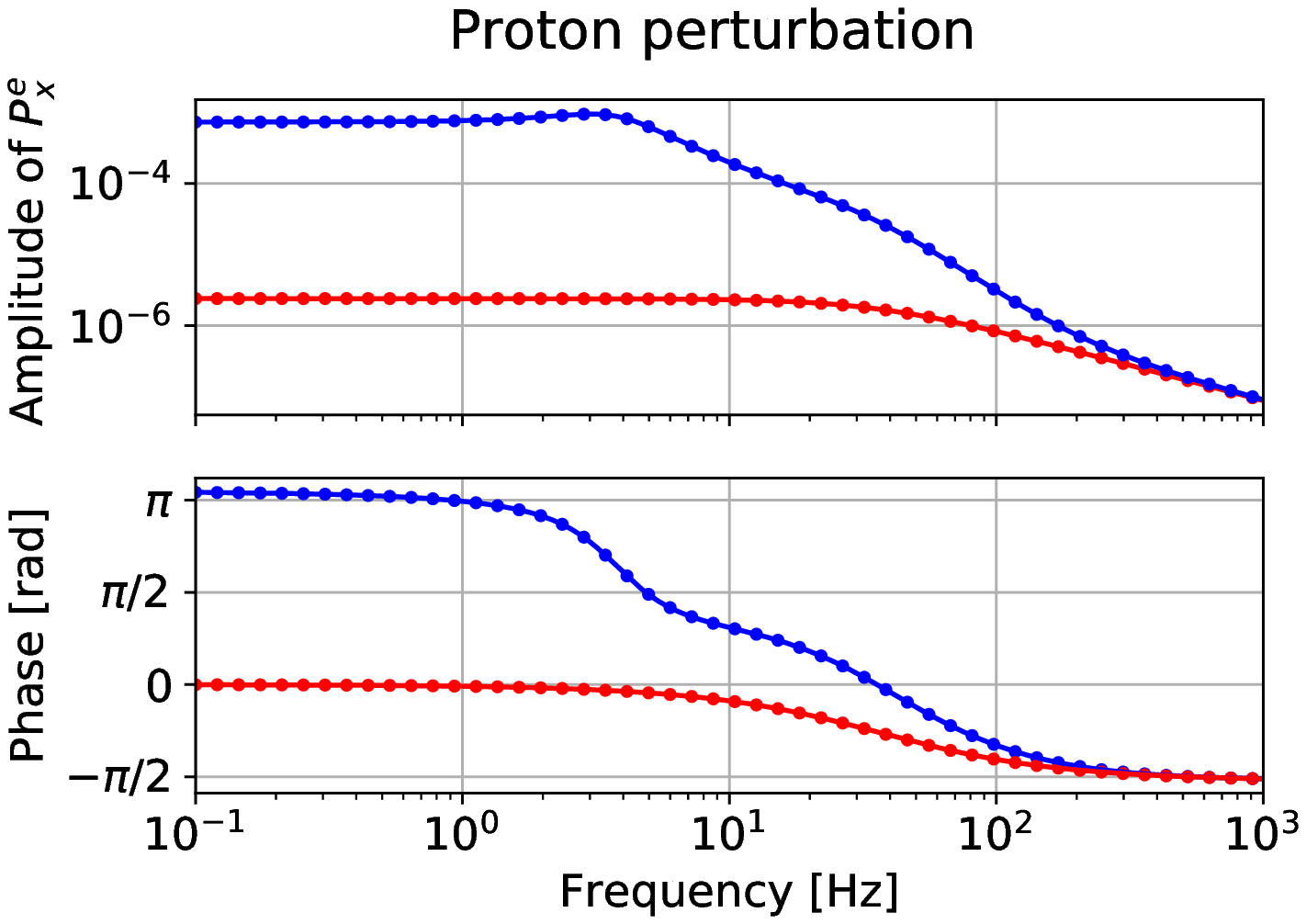}
    \end{center}
\end{minipage}
\vspace{0.3 cm}
 \caption{Numerically calculated (dots) and theoretically simulated (lines) amplitude and phase responses of the SERF (red) and co-magnetometer (blue) magnetic (a), electron nonmagnetic (b), neutron nonmagnetic (c), and proton nonmagnetic (d) perturbations.}
\label{fig:same_coupling_bandwidth}
\end{figure}

The fitted amplitude and phase of the signals arising due to magnetic, electron nonmagnetic, neutron nonmagnetic, and proton nonmagnetic perturbations are shown in Fig.~\ref{fig:same_coupling_bandwidth}. For  perturbation amplitudes $A_0$ small enough to provide a linear response of the system, the numerical results (points) are in good agreement with results of simulations performed within an analytical model. In our model, it was assumed that the AM and NG longitudinal components of the polarisation are not affected by the \old{pulse} \new{perturbation}, i.e., the longitudinal polarisations are constant over time, being equal to steady-state polarisation. For more details about analytical solutions see Additional information.

A key feature of the co-magnetometer in searches for pseudo-magnetic spin couplings is its suppressed sensitivity to low-frequency magnetic-field perturbations.  This is clearly visible in the data presented in Fig.~\ref{fig:same_coupling_bandwidth}(a).  At the lowest frequencies, the response amplitude of the co-magnetometer to the magnetic field is roughly four orders of magnitude lower than the response of the SERF magnetometer.  This difference decreases at higher frequencies reducing to zero at about 4~Hz.  Since compensation is provided by the NG, which adiabatically follows the field changes, and, at the compensation point, the atoms are only experiencing a field proportional to their own magnetisation, $B^i\approx B_c+\lambda M^{i'}P^{i'}_0=\lambda M^iP^i_0$, the frequency at which the SERF and co-magnetometer magnetic responses are equal is determined by the NG Larmor frequency.  Shifting the compensation point towards higher frequencies requires increasing the NG magnetisation, which can be achieved by  either  increasing the NG concentration or the polarisation. Both may be challenging experimentally.  For higher frequencies the response of both systems has the same amplitude, since outside of the self-compensating regime the co-magnetometer response is predominantly determined by the AM.  Thereby, the AM polarisation in the co-magnetometer starts to behave in the same way as the free-AM polarisation in a usual SERF magnetometer.   

There is a remarkable difference in the phase response of the two devices [Fig.~\ref{fig:same_coupling_bandwidth}(a)]. For magnetic field frequencies below the NG Larmor frequency (4~Hz), the response of the co-magnetometer is phase shifted by about $\pi$, and it  decreases sharply above that frequency, eventually becoming similar for both devices.  For lower frequencies, the difference is due to the NG that compensates the magnetic field and no such compensation is present in the SERF magnetometer.  For higher field frequencies the magnetic response of both devices is determined by the AM, so the observed dependencies are similar.

 When analysing the response of the co-magnetometer to nonmagnetic nuclear perturbation, it should be first noted that, unlike in the case of magnetic perturbation, the nonmagnetic nuclear perturbations are not compensated. Therefore, in that case, there is no reduction of the response amplitude at lower frequencies. To the contrary, the amplitude response to pseudo-magnetic nuclear perturbations of the co-magnetometer is significantly stronger than the response of the SERF magnetometer [Fig.~\ref{fig:same_coupling_bandwidth}(c)\&(d)].  In particular, for frequencies below 4~Hz, the co-magnetometer response is roughly five orders of magnitude stronger for the neutron nonmagnetic perturbation  [Fig.~\ref{fig:same_coupling_bandwidth}(c)], about three orders of magnitude stronger for the proton perturbation  [Fig.~\ref{fig:same_coupling_bandwidth}(d)] and even though it deteriorates for higher frequencies, it still remains significantly larger than for the SERF system. \new{Such dissimilarity  is related to the difference in effective pseudo-magnetic fields  experienced by NG and AM spins [see  Eqs.~\eqref{eq:sacaling_for_the_same_strength_num_value}].  Note, that the co-magnetometer has high sensitivity to the nuclear spin perturbations.   Specifically, because of the high concentration of the NG, the response of the co-magnetometer  is predominantly determined by the gas.} In turn, the large concentration difference between the NG and AM concentration (about six order of magnitude) is responsible for much higher sensitivity of the former to the nonmagnetic nuclear couplings.  Moreover, the high concentration and hence the high, compared to AM, magnetisation of the NG atoms (despite the 5\% polarisation of NG) ensures efficient transfer of the NG-spin perturbation to the AM polarisation. Therefore, the NG magnetisation-mediated nuclear coupling significantly increases the amplitude of the response of the co-magnetometer to nonmagnetic nuclear perturbations. 
An additional cause of the difference in the response to nuclear perturbations  stems from  the different nuclear spin contents of the $^{39}$K and the $^{3}$He nuclei \cite{Kimball_2015}. Contribution of the proton polarisation in $^{39}$K is roughly  four times larger than in the case of  $^{3}$He. In contrast,  the neutron polarisation has an about 24 times bigger contribution to the nuclear spin of $^{3}$He, than  it contributes to the nuclear spin of $^{39}$K. The difference in neutron and proton spin fractions in $^3$He nuclear spin also leads to different response amplitudes of the co-magnetometer to proton and neutron perturbations.

The phase response of the co-magnetometer is similar for both the neutron and proton pseudo-magnetic perturbations. Specifically, below {4~Hz}, the phase shift between perturbation and response is close to $\pi$ and it drops to about zero for higher frequencies. While for frequencies above 100~Hz the two responses differ, it should be noted that this frequency range is well beyond the bandwidth of the co-magnetometer, where amplitude of the response drops by several orders of magnitude. At the same time, the phase response of the SERF magnetometer is the same for magnetic and nuclear nonmagnetic perturbations, being zero at lower frequencies and monotonically shifting toward $-\pi/2$ for frequencies beyond the bandwidth of the magnetometer.

The  response of both devices to the electron nonmagnetic perturbation is similar at most of the frequencies with a distinct exception of the frequency corresponding to the NG Larmor frequency (4~Hz) [Fig.~\ref{fig:same_coupling_bandwidth}(b)].  
Such a behaviour is not surprising since, in both cases, the electron coupling perturbs the AM electron spins.  On the other hand, differences in the response of the systems at the NG Larmor frequency arise due to the coupling between the perturbed AM and the NG atoms.

\subsection*{Co-magnetometer response to transients }
It was shown in the previous section that the response of the co-magnetometer to nonmagnetic nuclear couplings is much stronger than that of the SERF magnetometer. Therefore, below we only focus on analysis of the response of the co-magnetometer to transient magnetic and nonmagnetic perturbations.  

\begin{figure}[h!]
    \centering
    \includegraphics[width = 0.5\linewidth]{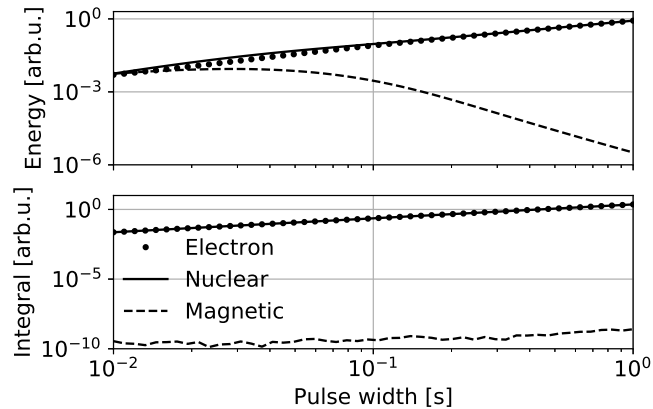}
    \caption{Energy and absolute value of the integral over temporal response of the co-magnetometer to the Lorentzian pulses of different width and coupling origin: electron nonmagnetic (dotted lines), nuclear nonmagnetic (solid lines), and magnetic (dashed lines) spin perturbations.  Presented results are based on numerical calculation of the 200-s long co-magnetometer responses to the different-origin pulse perturbations. The shape of the pulses is given by Eq.~\eqref{eq:lorentzian_pulse} and is centred at $t_0=100$~s, i.e., the centre of the simulation region. For all types of pulses, we assume the same amplitude given in the effective pseudo-magnetic field units.}
    \label{fig:final_result}
\end{figure}

As a generic example, we take a temporal Lorentzian perturbation of the amplitude $\Lambda_0$ and half-width $\Delta t$, centred at the time $t_0$, which is directed along the $\mathbf{y}$ axis (this can be easily generalised for any pulse shape)
\begin{equation}
    \bm{\Lambda_t} = \mathbf{y} \frac{\Lambda_0 \Delta t^2}{(t-t_0)^2 + \Delta t^2}.
    \label{eq:lorentzian_pulse}
\end{equation}
Such a definition allows keeping the amplitude of the pulse constant while varying its width (note that here the energy of the pulse is not preserved). Since here we are only interested in temporal parameters of the response, the proton and nuclear couplings are not considered independently but they are treated as a generic nuclear coupling. We simulated 200-s long responses of the co-magnetometer to the perturbations of different origin and the shape given by Eq.~\eqref{eq:lorentzian_pulse}, centred at $t_0 = 100$ s. We assumed the same amplitude of pulses in effective pseudo-magnetic field units for all types of the perturbation \new{to be small enough that the co-magnetometer continuously operates in self-compensating regime}. For all results we numerically calculated the integral over the response of the system and the energy of the signal within the simulated window. Results, presented in Fig.~\ref{fig:final_result}, are discussed below in  details.

\begin{figure}[h!]
\centering
\begin{minipage}{0.49\linewidth}
    \textbf{(a)}
    \begin{center}
    \includegraphics[width = 0.95\linewidth]{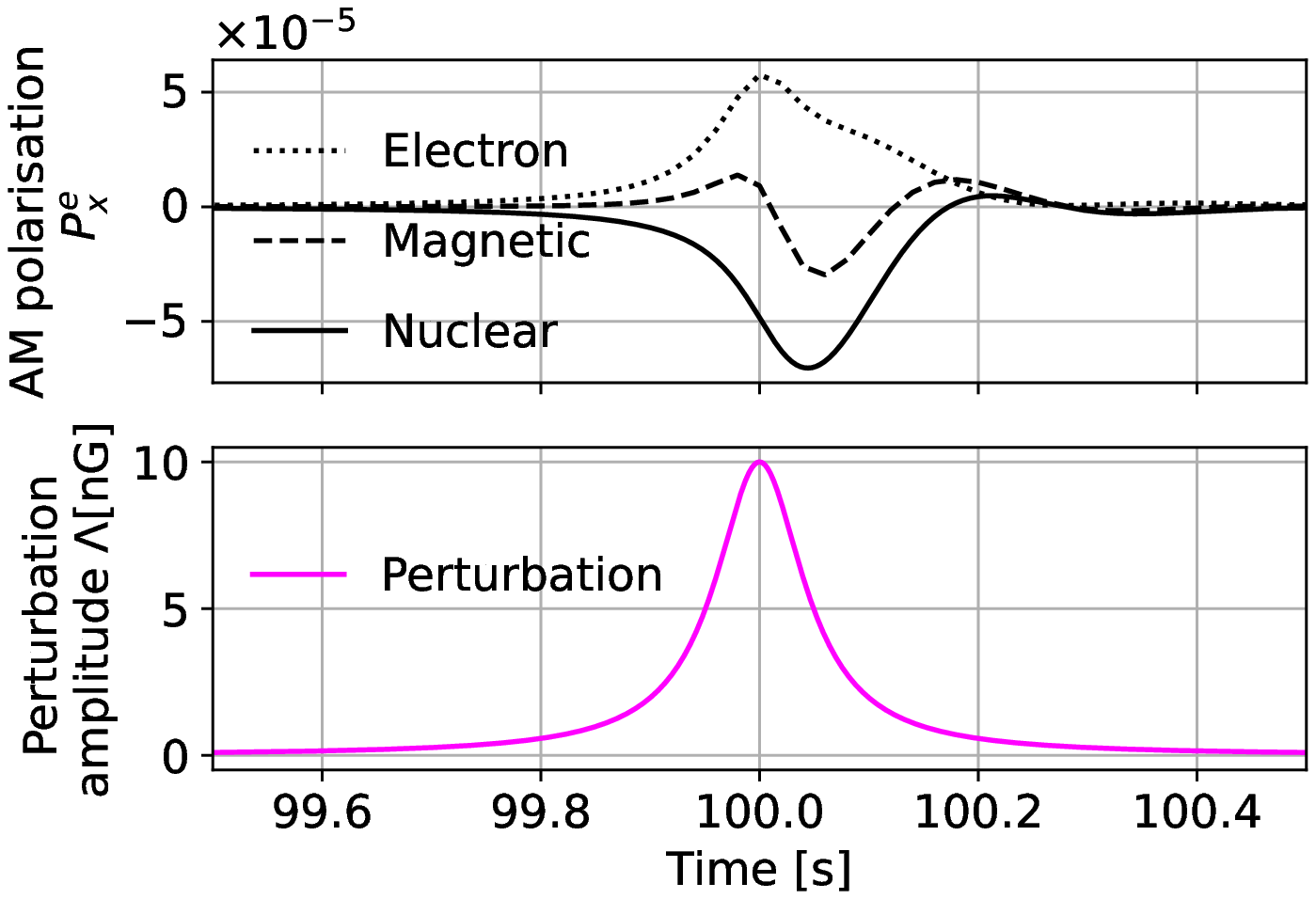}
    \end{center}
\end{minipage}
\hfil
\begin{minipage}{0.49\linewidth}
    \textbf{(b)}
    \begin{center}
    \includegraphics[width = 0.95\linewidth]{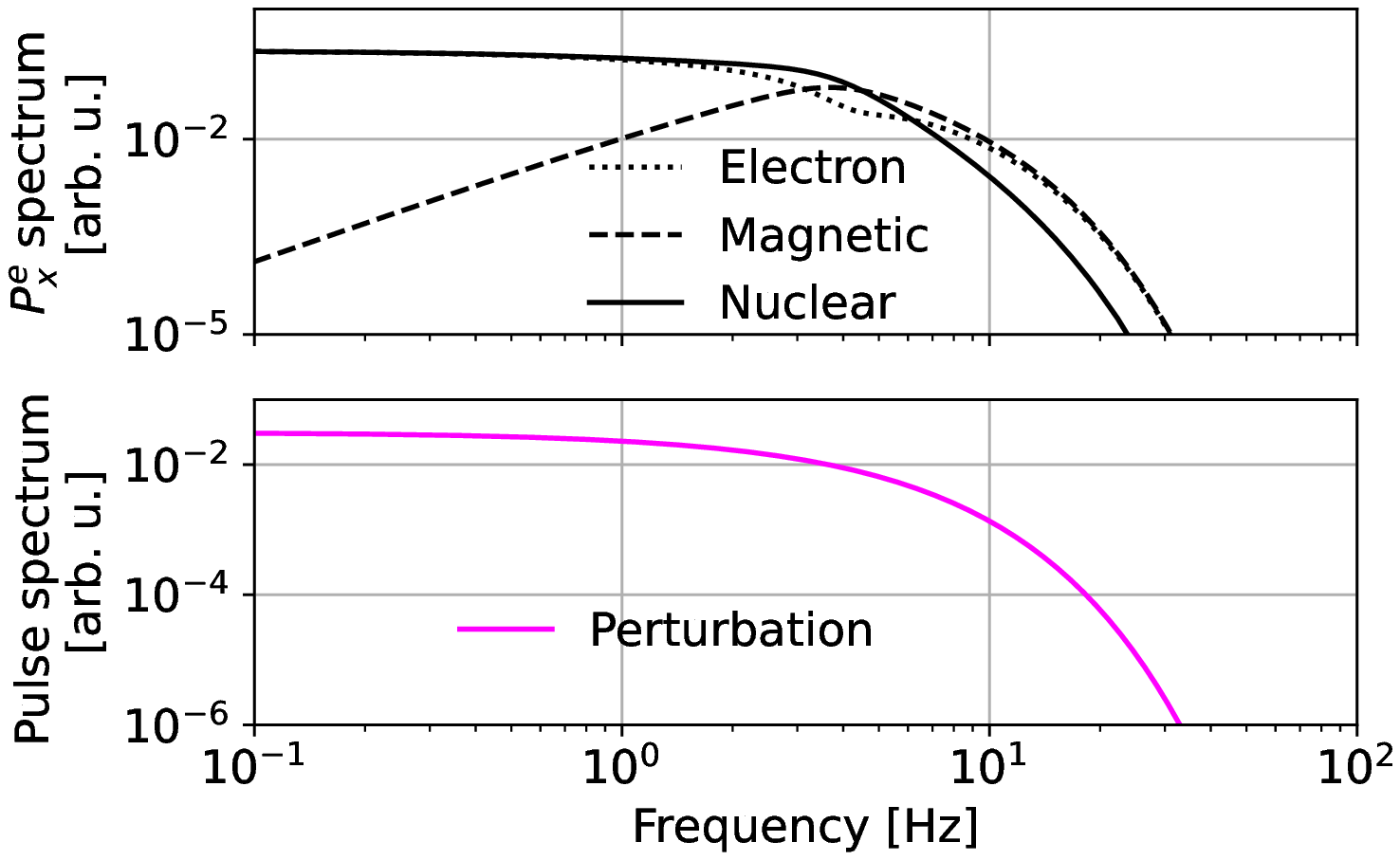}
    \end{center}
\end{minipage}

\caption{(a) Temporal responses of the co-magnetometer to pseudo-magnetic electron (dotted line), pseudo-magnetic nuclear (solid line) and magnetic (dashed line) spin perturbations. The bottom subplot shows the \new{50-ms  Lorentzian-shaped} perturbation common for all types of couplings. (b) \old{Frequency responses} \new{Spectra} of the co-magnetometer \new{responses} to pseudo-magnetic electron (dotted line), pseudo-magnetic nuclear (solid line) and magnetic (dashed line) spin perturbations  assuming the same amplitude of the perturbations in effective pseudo-magnetic magnetic field units. For the reference, the bottom subplot shows \old{spectra of the 1-s long (solid line) and} \new{spectrum of} 50-ms  \old{(dashed line)} Lorentzian pulses. \new{All spectra were obtained  with Fast Fourier Transform of 200-s long time series of the responses and perturbation.}}
\label{fig:pulses}
\end{figure}

In the previous section, we have shown that the co-magnetometer frequency responses for magnetic and nonmagnetic spin couplings are different. 
In particular, the compensation of low-frequency magnetic field \new{ in the co-magnetometer results in a suppression of low-frequency components of the pulse [Fig.~\ref{fig:pulses}(b)(top)] manifesting itself  in a response of the co-magnetometer to perturbations of different nature [Fig.~\ref{fig:pulses}(a)(top)].
The results show that the response significantly deviates from the Lorentzian shape of the magnetic pulse; the pulse is slightly longer and its shape is significantly distorted. At the same time, the distortion is much smaller both in the case of electron and nuclear perturbations, which is a manifestation of the absence of such compensation for pseudo-magnetic spin interaction.  Suppression of the low-frequency components leads to a weaker response of the device to the magnetic pulse with a spectrum  within the self-compensation band. This is shown in Fig.~\ref{fig:final_result}, where the pulse {integral and its energy are} presented versus the pulse length. On contrary, no such behaviour is observed  for nonmagnetic pulses, where low-frequency components are not suppressed [Fig.~\ref{fig:pulses}(b)(top)]. These results are confirmed with the simulations {for a wide range of pulse widths}; the longer the pulse, the more prominent is the divergence in energy between the response to magnetic and nonmagnetic perturbations [Fig.~\ref{fig:final_result}]. For 1-s long pulses, the difference is more than five orders of magnitude, and it grows for longer pulses. As for the pulse widths around 0.01 s, the response energy is comparable for all types of perturbations, since for the short pulses, a significant fraction of the pulse spectrum is at higher frequencies, where response is comparable and large for both magnetic and nonmagnetic perturbations.}

\new{The suppression of the low-frequency magnetic field in the co-magnetometer} provides another remarkable feature of the system; the integral over the co-magnetometer signal is significantly suppressed for magnetic field perturbations when integrated over time intervals significantly longer than the pulse width (for the presented results 200-s-long integration window have been used). In contrast, the integral is finite for nonmagnetic spin perturbations. For the simulated co-magnetometer system, assuming that the value of the effective pseudo-magnetic field is the same as the magnetic field, the difference between integrals over detected magnetic and nonmagnetic  transients is about seven orders of magnitude for pulse widths between 0.01 s to 1 s [Fig.~\ref{fig:final_result}].

\section*{Conclusions}
Numerical simulations of SERF and AM-NG co-magnetometers show a significantly stronger response of the latter to  nuclear nonmagnetic spin perturbations. 
While the  enhancement of the response due to the proton-coupling stems from  the high sensitivity of the co-magnetometer to the nuclear spin perturbations, 
 a larger contribution of the neutron polarisation to the NG polarisation provides an additional enhancement of the co-magnetometer response to neutron pseudo-magnetic perturbation. 
 At the same time, the response of both devices to the electron nonmagnetic spin perturbations is similar.
 
Our results demonstrate benefits of the co-magnetometer in searches for transient pseudo-magnetic spin couplings. On one hand, there is a suppressed response to low-frequency magnetic fields, which reduces noise of the device, on the other, due to ``high-pass'' magnetic-filter nature of the co-magnetometer, the device allows to differentiate between the magnetic and nonmagnetic transient responses, enabling a new way of identification of the observed signal nature. Specifically, the integral over time series signal for magnetic pulses has very small value, while it remains finite for nonmagnetic pulses.  \new{We also demonstrated that over a wide range of the pulse widths the energy of response to nonmagnetic perturbation is orders of magnitude greater than to magnetic-field perturbation. Even thought, the energy difference decreases for shorter pulse length, hence pulse energies, the difference in the integrals over the responses to nonmagnetic and magnetic perturbations remains large (several orders of magnitude) for the pulses which can be detected with the co-magnetometer.} The features of the co-magnetometer presented at this work demonstrate the capabilities of the co-magnetometer in searches for transient nonmagnetic spin couplings, i.e., the signals that are being searched by the GNOME.

\bibliography{sample}

\begin{thebibliography}{10}
\urlstyle{rm}
\expandafter\ifx\csname url\endcsname\relax
  \def\url#1{\texttt{#1}}\fi
\expandafter\ifx\csname urlprefix\endcsname\relax\def\urlprefix{URL }\fi
\expandafter\ifx\csname doiprefix\endcsname\relax\def\doiprefix{DOI: }\fi
\providecommand{\bibinfo}[2]{#2}
\providecommand{\eprint}[2][]{\url{#2}}

\bibitem{budker2013optical}
\bibinfo{editor}{Budker, D.} \& \bibinfo{editor}{Jackson~Kimball, D.} (eds.)
  \emph{\bibinfo{title}{Optical Magnetometry}} (\bibinfo{publisher}{Cambridge
  University Press}, \bibinfo{year}{2013}).

\bibitem{Safronova2018Search}
\bibinfo{author}{Safronova, M.~S.} \emph{et~al.}
\newblock \bibinfo{journal}{\bibinfo{title}{Search for new physics with atoms
  and molecules}}.
\newblock {\emph{\JournalTitle{Rev. Mod. Phys.}}}
  \textbf{\bibinfo{volume}{90}}, \bibinfo{pages}{025008},
  \doiprefix\url{10.1103/RevModPhys.90.025008} (\bibinfo{year}{2018}).

\bibitem{Kimball2018Searching}
\bibinfo{author}{{Jackson Kimball}, D.} \emph{et~al.}
\newblock \bibinfo{journal}{\bibinfo{title}{{Searching for axion stars and Q
  -balls with a terrestrial magnetometer network}}}.
\newblock {\emph{\JournalTitle{Physical Review D}}}
  \textbf{\bibinfo{volume}{97}}, \doiprefix\url{10.1103/PhysRevD.97.043002}
  (\bibinfo{year}{2018}).

\bibitem{Pospelov2013Detecting}
\bibinfo{author}{Pospelov, M.} \emph{et~al.}
\newblock \bibinfo{journal}{\bibinfo{title}{{Detecting domain walls of
  axionlike models using terrestrial experiments}}}.
\newblock {\emph{\JournalTitle{Physical Review Letters}}}
  \textbf{\bibinfo{volume}{110}},
  \doiprefix\url{10.1103/PhysRevLett.110.021803} (\bibinfo{year}{2013}).

\bibitem{Dailey2021Quantum}
\bibinfo{author}{Dailey, C.} \emph{et~al.}
\newblock \bibinfo{journal}{\bibinfo{title}{{Quantum sensor networks as exotic
  field telescopes for multi-messenger astronomy}}}.
\newblock {\emph{\JournalTitle{Nat. Astron.}}} \textbf{\bibinfo{volume}{5}},
  \doiprefix\url{10.1038/s41550-020-01242-7} (\bibinfo{year}{2021}).

\bibitem{Pustelny2013Global}
\bibinfo{author}{Pustelny, S.} \emph{et~al.}
\newblock \bibinfo{journal}{\bibinfo{title}{{The Global Network of Optical
  Magnetometers for Exotic physics (GNOME): A novel scheme to search for
  physics beyond the Standard Model}}}.
\newblock {\emph{\JournalTitle{Ann. der Phys.}}}
  \textbf{\bibinfo{volume}{525}}, \doiprefix\url{10.1002/andp.201300061}
  (\bibinfo{year}{2013}).

\bibitem{Afach2018Characterization}
\bibinfo{author}{Afach, S.} \emph{et~al.}
\newblock \bibinfo{journal}{\bibinfo{title}{Characterization of the global
  network of optical magnetometers to search for exotic physics ({GNOME})}}.
\newblock {\emph{\JournalTitle{Physics of the Dark Universe}}}
  \textbf{\bibinfo{volume}{22}}, \bibinfo{pages}{162--180},
  \doiprefix\url{10.1016/j.dark.2018.10.002} (\bibinfo{year}{2018}).

\bibitem{afach2021search}
\bibinfo{author}{Afach, S.} \emph{et~al.}
\newblock \bibinfo{title}{Search for topological defect dark matter using the
  global network of optical magnetometers for exotic physics searches
  ({GNOME})} (\bibinfo{year}{2021}).
\newblock \eprint{2102.13379}.

\bibitem{Kimball_2015}
\bibinfo{author}{Jackson~Kimball, D.~F.}
\newblock \bibinfo{journal}{\bibinfo{title}{Nuclear spin content and
  constraints on exotic spin-dependent couplings}}.
\newblock {\emph{\JournalTitle{New J. Phys.}}} \textbf{\bibinfo{volume}{17}},
  \bibinfo{pages}{073008}, \doiprefix\url{10.1088/1367-2630/17/7/073008}
  (\bibinfo{year}{2015}).

\bibitem{MASIAROIG2020_analysis_method}
\bibinfo{author}{Masia-Roig, H.} \emph{et~al.}
\newblock \bibinfo{journal}{\bibinfo{title}{Analysis method for detecting
  topological defect dark matter with a global magnetometer network}}.
\newblock {\emph{\JournalTitle{Physics of the Dark Universe}}}
  \textbf{\bibinfo{volume}{28}}, \bibinfo{pages}{100494},
  \doiprefix\url{https://doi.org/10.1016/j.dark.2020.100494}
  (\bibinfo{year}{2020}).

\bibitem{Romalis2002Dynamic}
\bibinfo{author}{Kornack, T.~W.} \& \bibinfo{author}{Romalis, M.~V.}
\newblock \bibinfo{journal}{\bibinfo{title}{Dynamics of two overlapping spin
  ensembles interacting by spin exchange}}.
\newblock {\emph{\JournalTitle{Phys. Rev. Lett.}}}
  \textbf{\bibinfo{volume}{89}}, \bibinfo{pages}{253002},
  \doiprefix\url{10.1103/PhysRevLett.89.253002} (\bibinfo{year}{2002}).

\bibitem{Kornack_Nuclear_spin_gyro}
\bibinfo{author}{Kornack, T.~W.}, \bibinfo{author}{Ghosh, R.~K.} \&
  \bibinfo{author}{Romalis, M.~V.}
\newblock \bibinfo{journal}{\bibinfo{title}{Nuclear spin gyroscope based on an
  atomic comagnetometer}}.
\newblock {\emph{\JournalTitle{Phys. Rev. Lett.}}}
  \textbf{\bibinfo{volume}{95}}, \bibinfo{pages}{230801},
  \doiprefix\url{10.1103/PhysRevLett.95.230801} (\bibinfo{year}{2005}).

\bibitem{Ghosh_Romalis_SEOP_Ne}
\bibinfo{author}{Ghosh, R.~K.} \& \bibinfo{author}{Romalis, M.~V.}
\newblock \bibinfo{journal}{\bibinfo{title}{Measurement of spin-exchange and
  relaxation parameters for polarizing $^{21}\mathrm{Ne}$ with {K} and {Rb}}}.
\newblock {\emph{\JournalTitle{Phys. Rev. A}}} \textbf{\bibinfo{volume}{81}},
  \bibinfo{pages}{043415}, \doiprefix\url{10.1103/PhysRevA.81.043415}
  (\bibinfo{year}{2010}).

\bibitem{Rot_sens_Li}
\bibinfo{author}{Li, R.} \emph{et~al.}
\newblock \bibinfo{journal}{\bibinfo{title}{Rotation sensing using a
  k-rb-$^{21}\mathrm{Ne}$ comagnetometer}}.
\newblock {\emph{\JournalTitle{Phys. Rev. A}}} \textbf{\bibinfo{volume}{94}},
  \bibinfo{pages}{032109}, \doiprefix\url{10.1103/PhysRevA.94.032109}
  (\bibinfo{year}{2016}).

\bibitem{fang2016low}
\bibinfo{author}{Fang, J.} \emph{et~al.}
\newblock \bibinfo{journal}{\bibinfo{title}{Low frequency magnetic field
  suppression in an atomic spin co-magnetometer with a large electron magnetic
  field}}.
\newblock {\emph{\JournalTitle{Journal of Physics B: Atomic, Molecular and
  Optical Physics}}} \textbf{\bibinfo{volume}{49}}, \bibinfo{pages}{065006}
  (\bibinfo{year}{2016}).

\bibitem{chen2016spin}
\bibinfo{author}{Chen, Y.} \emph{et~al.}
\newblock \bibinfo{journal}{\bibinfo{title}{Spin exchange broadening of
  magnetic resonance lines in a high-sensitivity rotating {K}-{Rb}-$^{21}${Ne}
  co-magnetometer}}.
\newblock {\emph{\JournalTitle{Sci. Rep.}}} \textbf{\bibinfo{volume}{6}},
  \bibinfo{pages}{1--12} (\bibinfo{year}{2016}).

\bibitem{Co-mag_magn_field_resp_Fan}
\bibinfo{author}{{Fan}, W.}, \bibinfo{author}{{Quan}, W.},
  \bibinfo{author}{{Zhang}, W.}, \bibinfo{author}{{Xing}, L.} \&
  \bibinfo{author}{{Liu}, G.}
\newblock \bibinfo{journal}{\bibinfo{title}{Analysis on the magnetic field
  response for nuclear spin co-magnetometer operated in spin-exchange
  relaxation-free regime}}.
\newblock {\emph{\JournalTitle{IEEE Access}}} \textbf{\bibinfo{volume}{7}},
  \bibinfo{pages}{28574--28580}, \doiprefix\url{10.1109/ACCESS.2019.2902181}
  (\bibinfo{year}{2019}).

\bibitem{Shi:20}
\bibinfo{author}{Shi, M.}
\newblock \bibinfo{journal}{\bibinfo{title}{Investigation on magnetic field
  response of a $^{87}${Rb}-$^{129}${Xe} atomic spin comagnetometer}}.
\newblock {\emph{\JournalTitle{Opt. Express}}} \textbf{\bibinfo{volume}{28}},
  \bibinfo{pages}{32033--32041}, \doiprefix\url{10.1364/OE.404809}
  (\bibinfo{year}{2020}).

\bibitem{theory_of_he_SEOP}
\bibinfo{author}{Appelt, S.} \emph{et~al.}
\newblock \bibinfo{journal}{\bibinfo{title}{Theory of spin-exchange optical
  pumping of ${}^{3}\mathrm{He}$ and ${}^{129}\mathrm{Xe}$}}.
\newblock {\emph{\JournalTitle{Phys. Rev. A}}} \textbf{\bibinfo{volume}{58}},
  \bibinfo{pages}{1412--1439}, \doiprefix\url{10.1103/PhysRevA.58.1412}
  (\bibinfo{year}{1998}).

\bibitem{allred2002high}
\bibinfo{author}{Allred, J.}, \bibinfo{author}{Lyman, R.},
  \bibinfo{author}{Kornack, T.} \& \bibinfo{author}{Romalis, M.~V.}
\newblock \bibinfo{journal}{\bibinfo{title}{High-sensitivity atomic
  magnetometer unaffected by spin-exchange relaxation}}.
\newblock {\emph{\JournalTitle{Phys. Rev; Lett.}}}
  \textbf{\bibinfo{volume}{89}}, \bibinfo{pages}{130801}
  (\bibinfo{year}{2002}).

\bibitem{Kornack_phdthesis}
\bibinfo{author}{Kornack, T.}
\newblock \emph{\bibinfo{title}{A test of {CPT} and {Lorentz} symmetry using a
  {K}-$^3${He} co-magnetometer}}.
\newblock Ph.D. thesis, \bibinfo{school}{Princeton University}
  (\bibinfo{year}{2005}).

\bibitem{new_CPT_limit}
\bibinfo{author}{Brown, J.~M.}, \bibinfo{author}{Smullin, S.~J.},
  \bibinfo{author}{Kornack, T.~W.} \& \bibinfo{author}{Romalis, M.~V.}
\newblock \bibinfo{journal}{\bibinfo{title}{New limit on {Lorentz}- and
  {CPT}-violating neutron spin interactions}}.
\newblock {\emph{\JournalTitle{Phys. Rev. Lett.}}}
  \textbf{\bibinfo{volume}{105}}, \bibinfo{pages}{151604},
  \doiprefix\url{10.1103/PhysRevLett.105.151604} (\bibinfo{year}{2010}).

\bibitem{New_test_of_local_lorentz_invariance}
\bibinfo{author}{Smiciklas, M.}, \bibinfo{author}{Brown, J.~M.},
  \bibinfo{author}{Cheuk, L.~W.}, \bibinfo{author}{Smullin, S.~J.} \&
  \bibinfo{author}{Romalis, M.~V.}
\newblock \bibinfo{journal}{\bibinfo{title}{New test of local {Lorentz}
  invariance using a $^{21}${Ne}-{Rb}-{K} comagnetometer}}.
\newblock {\emph{\JournalTitle{Phys. Rev. Lett.}}}
  \textbf{\bibinfo{volume}{107}}, \bibinfo{pages}{171604},
  \doiprefix\url{10.1103/PhysRevLett.107.171604} (\bibinfo{year}{2011}).

\bibitem{Limits_on_new_long_range_nuclear_spin_dep_forces_Romaslis}
\bibinfo{author}{Vasilakis, G.}, \bibinfo{author}{Brown, J.~M.},
  \bibinfo{author}{Kornack, T.~W.} \& \bibinfo{author}{Romalis, M.~V.}
\newblock \bibinfo{journal}{\bibinfo{title}{Limits on new long range nuclear
  spin-dependent forces set with a {K}-$^3${He} comagnetometer}}.
\newblock {\emph{\JournalTitle{Phys. Rev. Lett.}}}
  \textbf{\bibinfo{volume}{103}}, \bibinfo{pages}{261801},
  \doiprefix\url{10.1103/PhysRevLett.103.261801} (\bibinfo{year}{2009}).

\bibitem{New_limit_spin_mass_lee}
\bibinfo{author}{Lee, J.}, \bibinfo{author}{Almasi, A.} \&
  \bibinfo{author}{Romalis, M.}
\newblock \bibinfo{journal}{\bibinfo{title}{Improved limits on spin-mass
  interactions}}.
\newblock {\emph{\JournalTitle{Phys. Rev. Lett.}}}
  \textbf{\bibinfo{volume}{120}}, \bibinfo{pages}{161801},
  \doiprefix\url{10.1103/PhysRevLett.120.161801} (\bibinfo{year}{2018}).

\bibitem{Romalis_Savukov_2005}
\bibinfo{author}{Savukov, I.~M.} \& \bibinfo{author}{Romalis, M.~V.}
\newblock \bibinfo{journal}{\bibinfo{title}{Effects of spin-exchange collisions
  in a high-density alkali-metal vapor in low magnetic fields}}.
\newblock {\emph{\JournalTitle{Phys. Rev. A}}} \textbf{\bibinfo{volume}{71}},
  \bibinfo{pages}{023405}, \doiprefix\url{10.1103/PhysRevA.71.023405}
  (\bibinfo{year}{2005}).

\bibitem{Schaefer_freq_shifts}
\bibinfo{author}{Schaefer, S.~R.} \emph{et~al.}
\newblock \bibinfo{journal}{\bibinfo{title}{Frequency shifts of the
  magnetic-resonance spectrum of mixtures of nuclear spin-polarized noble gases
  and vapors of spin-polarized alkali-metal atoms}}.
\newblock {\emph{\JournalTitle{Phys. Rev. A}}} \textbf{\bibinfo{volume}{39}},
  \bibinfo{pages}{5613--5623}, \doiprefix\url{10.1103/PhysRevA.39.5613}
  (\bibinfo{year}{1989}).

\end{thebibliography}

\section*{Acknowledgements}
The authors thank Derek Jackson Kimball for useful discussions. Preliminary work in the directions discussed in the manuscript was carried out by Samer Afach. The work of MP and MK was supported by the Jagiellonian University in Krakow, Faculty of Physics Astronomy and Applied Computer Science internal grant (N17/MNS/000007/2020). SP would like to acknowledge support of the National Science Centre of Poland within the OPUS programme. The work of DB was supported by the Cluster of Excellence ``Precision Physics, Fundamental Interactions, and Structure of Matter'' (PRISMA+ EXC 2118/1) funded by the German Research Foundation (DFG) within the German Excellence Strategy (Project ID 39083149), by the European Research Council (ERC) under the European Union Horizon 2020 research and innovation program (project Dark-OST, grant agreement No 695405), and by the DFG Reinhart Koselleck project. The work of AW was supported by  the German Federal Ministry of Education and Research (BMBF) within the Quantumtechnologien program (Grant No. 13N15064).

\section*{Author contributions statement}
M.P. defined a numerical models, conducted simulations, analysed the results, constructed analytical solution and wrote the manuscript, M.K. participated into definition of a numerical models and edited manuscript,  S.P. conceived and supervised the project, D.B., R.C., S.P., and A.W. analysed the results, discussed their interpretation and edited the manuscript. All authors reviewed the manuscript. 

\end{document}